\documentclass{iau}

\usepackage{stmaryrd}
\usepackage{graphicx}
\usepackage{natbib}
\usepackage{hyperref}
\hypersetup{
     breaklinks   = true,
     colorlinks   = true,
     citecolor    = blue,
     linkcolor    = blue,
     urlcolor    = blue
}

\newcommand{\apj}{ApJ} % The Astrophysical Journal
\newcommand{\apjl}{ApJ} % The Astrophysical Journal LettersWardleNg1999
\newcommand{\apjs}{ApJ} % The Astrophysical Journal Supplement
\newcommand{\aj}{AJ} % The Astronomical Journal
\newcommand{\aap}{A\&A} % Astronomy & Astrophysics
\newcommand{\araa}{ARAA} % Annual Review of Astronomy and Astrophysics
\newcommand{\mnras}{MNRAS} % Monthly Notices of the Royal Astronomical Society
\newcommand\aapr{A\&A~Rev.} % Astronomy and Astrophysics Reviews
\newcommand\ssr{SSRv} % Space Science Reviews
\title[IAUS 322.~~The CMZ cloud G0.253+0.016] %% give here short title %%
{The link between solenoidal turbulence and slow star formation in G0.253+0.016}

\author[Federrath et al.]   %% give here short author list %%
{C.~Federrath$^1$, J.~M.~Rathborne$^2$, S.~N.~Longmore$^3$, J.~M.~D.~Kruijssen$^{4,5}$, J.~Bally$^{6}$, Y.~Contreras$^{7}$, R.~M.~Crocker$^1$, G.~Garay$^8$, J.~M.~Jackson$^{9}$, L.~Testi$^{10,11,12}$, A.~J.~Walsh$^{13}$}

\affiliation{
$^1$Research School of Astronomy and Astrophysics, Australian National University, Canberra, ACT~2611, Australia \\ email: {\href{mailto:christoph.federrath@anu.edu.au}{\tt christoph.federrath@anu.edu.au}} \\[\affilskip]
$^2$CSIRO Astronomy and Space Science, P.O.~Box~76, Epping NSW, 1710, Australia \\[\affilskip]
$^3$Astrophysics Research Institute, Liverpool John Moores University, IC2, Liverpool Science Park, 146~Brownlow Hill, Liverpool~L3~5RF, United Kingdom \\[\affilskip]
$^4$Astronomisches Rechen-Institut, Zentrum f\"ur Astronomie der Universit\"at Heidelberg, M\"onchhofstra{\ss}e 12-14, 69120~Heidelberg, Germany \\[\affilskip]
$^5$Max-Planck Institut f\"{u}r Astronomie, K\"{o}nigstuhl~17, 69117~Heidelberg, Germany \\[\affilskip]
$^6$CASA, University of Colorado, 389-UCB, Boulder, CO~80309, USA \\[\affilskip]
$^7$Leiden Observatory, Leiden University, PO~Box~9513, NL-2300 RA Leiden, the Netherlands \\[\affilskip]
$^8$Departamento de Astronom\'ia, Universidad de Chile, Casilla 36-D, Santiago, Chile \\[\affilskip]
$^9$Institute for Astrophysical Research, Boston University, Boston, MA~02215, USA \\[\affilskip]
$^{10}$European Southern Observatory, Karl-Schwarzschild-Stra{\ss}e~2, D-85748 Garching bei M\"unchen, Germany \\[\affilskip]
$^{11}$INAF-Arcetri, Largo E.~Fermi~5, I-50125 Firenze, Italy \\[\affilskip]
$^{12}$Excellence Cluster Universe, Boltzmannstra{\ss}e~2, D-85748, Garching, Germany \\[\affilskip]
$^{13}$International Centre for Radio Astronomy Research, Curtin University, GPO Box U1987, Perth WA~6845, Australia
}

\pubyear{2016}
\volume{322}  %% insert here IAU Symposium No.
\jname{The Multi-Messenger Astrophysics of the Galactic Centre}
\editors{R.~M.~Crocker, S.~N.~Longmore \& G.~V.~Bicknell, eds.}
\begin{document}

\maketitle

% 150 words
\begin{abstract}
Star formation in the Galactic disc is primarily controlled by gravity, turbulence, and magnetic fields. It is not clear that this also applies to star formation near the Galactic Centre. Here we determine the turbulence and star formation in the CMZ cloud G0.253+0.016. Using maps of $3\,$mm dust emission and HNCO intensity-weighted velocity obtained with ALMA, we measure the volume-density variance $\sigma_{\rho/\rho_0}=1.3\pm0.5$ and turbulent Mach number $\mathcal{M}=11\pm3$. Combining these with turbulence simulations to constrain the plasma $\beta=0.34\pm0.35$, we reconstruct the turbulence driving parameter $b=0.22\pm0.12$ in G0.253+0.016. This low value of $b$ indicates solenoidal (divergence-free) driving of the turbulence in G0.253+0.016. By contrast, typical clouds in the Milky Way disc and spiral arms have a significant compressive (curl-free) driving component ($b>0.4$). We speculate that shear causes the solenoidal driving in G0.253+0.016 and show that this may reduce the star formation rate by a factor of $7$ compared to nearby clouds.
\keywords{Galaxy: centre, ISM: clouds, magnetic fields, stars: formation, turbulence}
\end{abstract}

\firstsection

\section{Introduction}

\citet{RathborneEtAl2014,RathborneEtAl2015} showed that {G0.253+0.016} is a dense turbulent cloud in the central molecular zone (CMZ). However, so far it has been unclear what drives this turbulence and whether that turbulence plays a role in controlling the low star formation rate (SFR) seen in {G0.253+0.016} and in the CMZ as a whole \citep{LongmoreEtAl2013a,KruijssenEtAl2014,JohnstonEtAl2014}. Using high-resolution ALMA $3\,$mm dust and HNCO molecular line data, we determine the driving mode of the turbulence in {G0.253+0.016} and link the turbulence driving to the SFR.

\begin{figure*}
\centerline{\includegraphics[width=1.0\linewidth]{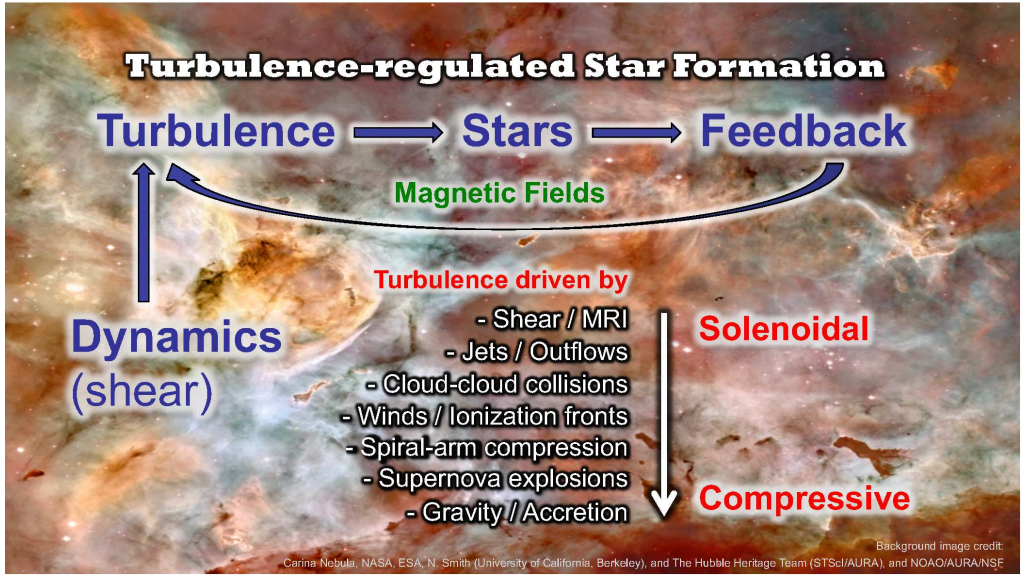}}
\caption{Sketch of the turbulence-regulated paradigm of star formation. Turbulence is fed by stellar feedback and/or large-scale dynamics (such as galactic shear). Different turbulence driving mechanisms can excite more solenoidal (rotational) modes, others inject more compressive (potential) modes. The mix of turbulent modes has profound consequences for star formation.}
\label{fig:turbsf}
\end{figure*}

The turbulence-regulated paradigm of star formation \citep{MacLowKlessen2004,ElmegreenScalo2004,McKeeOstriker2007,HennebelleFalgarone2012,FederrathKlessen2012,PadoanEtAl2014} provides us with the basic framework for our approach to determine the turbulence parameters of {G0.253+0.016} and allows us to make predictions for the star formation activity in {G0.253+0.016}. Figure~\ref{fig:turbsf} shows a sketch of the turbulence-regulated picture of star formation. In this model, turbulence shapes the density distribution of the clouds, thereby controlling the dense-gas fraction and thus, the formation of stars. Then, stellar feedback (such as supernova explosions or stellar winds) and/or large-scale dynamics (such as galactic shear or magneto-rotational instability) drive the turbulence. Understanding and determining the drivers of the turbulence is of fundamental importance in this model of star formation.

Idealised numerical simulations have shown that compressible, supersonic turbulence decays quickly, in about a crossing time \citep{ScaloPumphrey1982,MacLowEtAl1998,StoneOstrikerGammie1998,MacLow1999}. Given that {G0.253+0.016} and other galactic clouds are in a dynamic state of supersonic turbulence means that the turbulence is driven by some physical stirring mechanism(s).

Turbulence driving mechanisms can be broadly separated into two groups: 1) stellar feedback, and 2) gas dynamics caused by mechanisms other than feedback. Stellar feedback includes supernova explosions, stellar winds, and ionisation fronts \citep{McKee1989,KrumholzMatznerMcKee2006,BalsaraEtAl2004,AvillezBreitschwerdt2005,BreitschwerdtEtAl2009,GritschnederEtAl2009,PetersEtAl2010,PetersEtAl2011,ArceEtAl2011,GoldbaumEtAl2011,LeeMurrayRahman2012}, primarily caused by high-mass stars, as well as jets and outflows from young stars, including low- and intermediate-mass stars \citep{NormanSilk1980,MatznerMcKee2000,BanerjeeKlessenFendt2007,NakamuraLi2008,CunninghamEtAl2009,CarrollFrankBlackman2010,WangEtAl2010,CunninghamEtAl2011,PlunkettEtAl2013,PlunkettEtAl2015,OffnerArce2014,FederrathEtAl2014}. The 2nd category (which we refer to as ``Dynamics'' in Figure~\ref{fig:turbsf}) includes accretion (such as accretion onto a galaxy) and gravitational collapse \citep{Hoyle1953,VazquezCantoLizano1998,KlessenHennebelle2010,ElmegreenBurkert2010,VazquezSemadeniEtAl2010,FederrathSurSchleicherBanerjeeKlessen2011,RobertsonGoldreich2012,LeeChangMurray2015}, the magneto-rotational instability (MRI) \citep{BalbusHawley1991,PiontekOstriker2004,PiontekOstriker2007,TamburroEtAl2009}, spiral-arm compression \citep{DobbsBonnell2008,DobbsEtAl2008}, cloud-cloud collisions \citep{TaskerTan2009,BenincasaEtAl2013}, and shear. While different drivers can play a role in different environments, \citet{KruijssenEtAl2014} found that most of these drivers are not sufficient to explain the turbulent velocity dispersion in the CMZ, but some of them can.

A critical consideration is that the majority of turbulence drivers (e.g., supernova explosions, high-mass stellar winds, and accretion) primarily drive compressible (curl-free) modes, so we refer to these as ``compressive drivers''. By contrast, solenoidal (divergence-free) modes can be generated directly by shear and the MRI (so we call them ``solenoidal drivers''). The key aspect here is that the density probability distribution function (PDF) depends critically on the driving. \citet{FederrathKlessenSchmidt2008,FederrathDuvalKlessenSchmidtMacLow2010,PriceFederrathBrunt2011,MolinaEtAl2012,KonstandinEtAl2012ApJ,NolanFederrathSutherland2015,FederrathBanerjee2015} showed that the variance (width) of the density PDF is given by
\begin{equation} \label{eq:sigrho}
\sigma_{\rho/\rho_0} = b\,\mathcal{M}\left(1+\beta^{-1}\right)^{-1/2},
\end{equation}
with the turbulent Mach number $\mathcal{M}=\sigma_v/c_\mathrm{s}$ (the ratio of velocity dispersion and sound speed), plasma $\beta$ (the ratio of thermal and magnetic pressure), and the turbulence driving parameter $b$, which smoothly varies from $b=1/3$ for purely solenoidal driving to $b=1$ for purely compressive driving \citep{FederrathDuvalKlessenSchmidtMacLow2010}.

The theoretical models and simulations in \citet{FederrathKlessen2012} demonstrated that the SFR depends on $b$, with compressive driving producing up to an order of magnitude higher SFRs than solenoidal driving. Thus, our goal is to determine whether the driving of turbulence in {G0.253+0.016} is primarily solenoidal or compressive. We do this by measuring $\sigma_{\rho/\rho_0}$, $\mathcal{M}$, and $\beta$, and inverting Equation~(\ref{eq:sigrho}) to solve for $b$. Finally, we use our measurement of $b$ to predict the SFR in {G0.253+0.016} and to contrast this to the SFR in Milky Way clouds located in the Galactic disc rather than the Galactic Centre.

\section{Results}

The main results and methods of this work are published in \citet{FederrathEtAl2016}. Here we summarise the main results concerning the driving mode of the turbulence in {G0.253+0.016} and its implications for the SFR.

\begin{figure*}
\centerline{\includegraphics[width=1.0\linewidth]{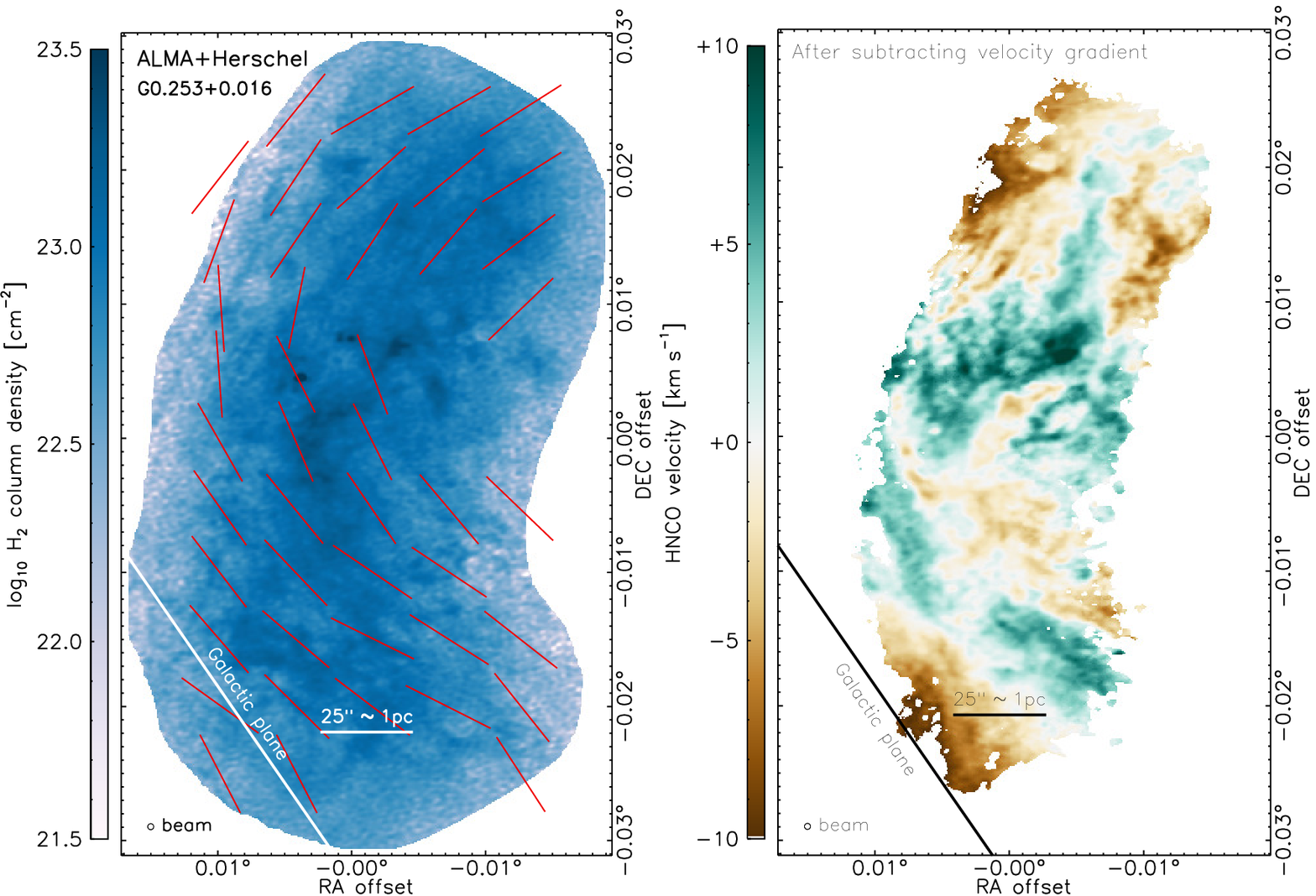}}
\caption{Left-hand panel: H$_2$ column density map of {G0.253+0.016} from combined ALMA+\emph{Herschel} data \citep{RathborneEtAl2014}. The direction of the large-scale magnetic field from polarisation measurements obtained in \citet{DotsonEtAl2010} and \citet{PillaiEtAl2015} is shown as red pseudo vectors. Right-hand panel: HNCO intensity-weighted velocity of {G0.253+0.016} after subtraction of the large-scale velocity gradient across the cloud. The gradient-subtracted map primarily depicts the turbulent motions in the plane of the sky. Both maps are in equatorial coordinates with the (0,0) offset position in right ascension (RA)~and declination (DEC) being 17:46:09.59, $-$28:42:34.2~J2000, respectively.}
\label{fig:images}
\end{figure*}

\begin{figure*}
\centerline{\includegraphics[width=1.0\linewidth]{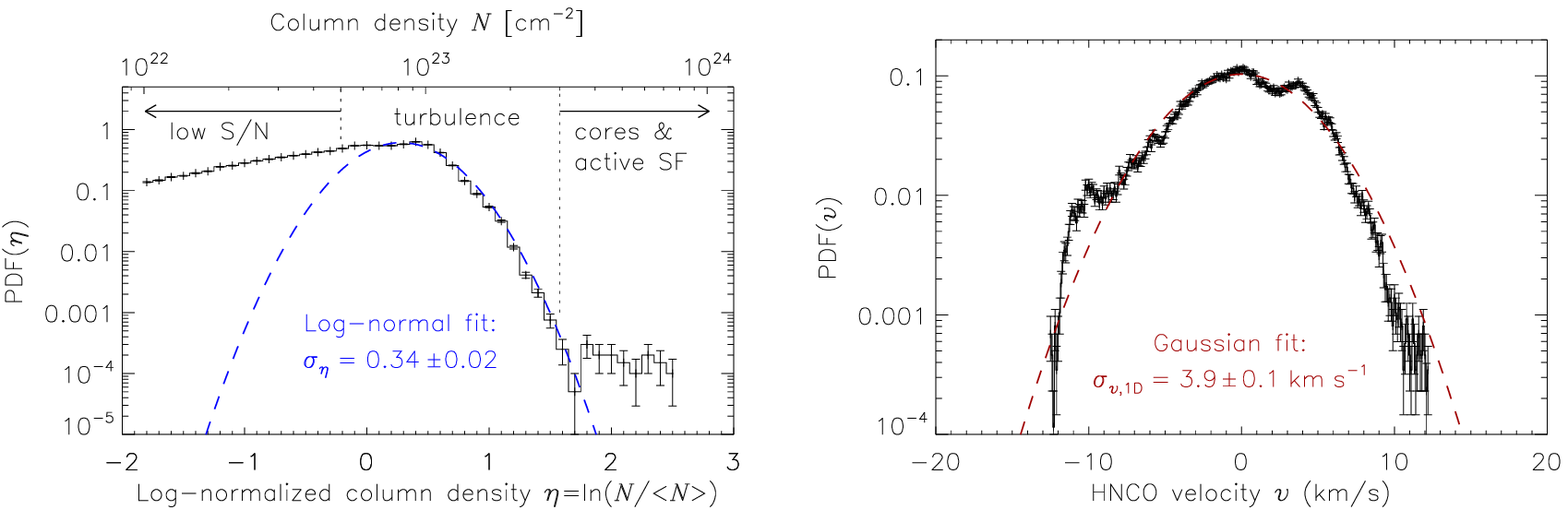}}
\caption{Left-hand panel: column density PDF of the map shown in the left-hand panel of Figure~\ref{fig:images}, for the log-normalised column density contrast $\eta=\ln(N/\langle N\rangle)$. A log-normal fit (dashed line) in the turbulence-dominated regime gives $\sigma_\eta = 0.34\pm0.02$ \citep{RathborneEtAl2014}. Right-hand panel: HNCO velocity PDF of the map shown in the right-hand panel of Figure~\ref{fig:images}. From a Gaussian fit, we measure the 1D turbulent velocity dispersion $\sigma_{v,\mathrm{1D}}=3.9\pm0.1\,\mathrm{km}\,\mathrm{s}^{-1}$.}
\label{fig:pdfs}
\end{figure*}

Figure~\ref{fig:images} shows the H$_2$ column density and the HNCO intensity-weighted velocity in {G0.253+0.016}. In order to isolate primarily turbulent motions in the cloud, we have subtracted the large-scale velocity gradient seen across {G0.253+0.016} \citep{RathborneEtAl2015}. From the two maps in Figure~\ref{fig:images}, we compute the column density PDF and the turbulent velocity PDF shown in Figure~\ref{fig:pdfs}, respectively. We find $\sigma_\eta=0.34\pm0.02$ \citep[as reported by][]{RathborneEtAl2014} for the log-normalised column density contrast $\eta=\ln(N/\langle N\rangle)$ and the 1D velocity dispersion $\sigma_{v,\mathrm{1D}}=3.9\pm0.1\,\mathrm{km}\,\mathrm{s}^{-1}$, both in the plane of the sky. Using temperature measurements from \citet{GinsburgEtAl2016}, we find sound speeds $c_\mathrm{s}=0.60\pm0.15\,\mathrm{km}\,\mathrm{s}^{-1}$, leading to the 3D turbulent Mach number, $\mathcal{M}=11\pm3$ in {G0.253+0.016}.

In order to apply Equation~(\ref{eq:sigrho}) to solve for the turbulence driving parameter $b$, we need to convert the column-density variance $\sigma_\eta$ to the volume-density variance $\sigma_{\rho/\rho_0}$. We use the technique developed in \citet{BruntFederrathPrice2010a,BruntFederrathPrice2010b} \citep[see also][]{KainulainenFederrathHenning2014} and find $\sigma_{\rho/\rho_0}=1.3\pm0.5$. Finally, we need an estimate for the turbulent (unordered) magnetic-field plasma $\beta$ parameter. We use the ordered magnetic field measurement from \citet{PillaiEtAl2015} (see polarisation pseudo vectors in Figure~\ref{fig:images}) and run magneto-hydrodynamical turbulence simulations with this ordered field and the measured $\mathcal{M}$. From the simulations, we determine the unordered, turbulent plasma $\beta=0.34\pm0.35$ entering Equation~({\ref{eq:sigrho}}).

Combining our measurements of $\sigma_{\rho/\rho_0}$, $\mathcal{M}$, and $\beta$ and propagating the uncertainties, we find $b=0.22\pm0.12$ in {G0.253+0.016}, indicating solenoidal driving of the turbulence. We interpret this driving to be the result of strong shearing motions in the CMZ \citep[][Kruijssen et al., in prep.]{KrumholzKruijssen2015}. This is in stark contrast to the driving parameter inferred for typical clouds in the Galactic disc and spiral arms of the Milky Way. Currently available measurements for Taurus \citep{Brunt2010}, GRSMC43.30-0.33 \citep{GinsburgFederrathDarling2013} and IC5146 \citep{PadoanJonesNordlund1997} all show $b>0.4$ with typical values of $b=0.5$, thus exhibiting a significant compressive driving component, in contrast to the CMZ cloud {G0.253+0.016}.

\section{Implications for the SFR}
Using our measurements of $b$, $\mathcal{M}$, and $\beta$, we find a predicted $\mathrm{SFR}=(1.1\pm0.8)\times10^{-2}\,\mbox{$M_{\varodot}$}\,\mathrm{yr}^{-1}$, based on the theoretical multi-freefall models of the SFR summarised in \citet{FederrathKlessen2012}. If we used a turbulence driving parameter $b=0.5$ (as measured for clouds in the Galactic disc and spiral arms), we would find a 7 times higher SFR. This shows that the driving of the turbulence is a critical parameter for star formation and may significantly contribute to explaining the low SFRs in {G0.253+0.016} and possibly in the entire CMZ.

%\bibliographystyle{/Users/chfeder/Documents/Latex-sources/apj}
%\bibliography{/Users/chfeder/Documents/Latex-sources/federrath.bib}

%\begin{discussion}
%\end{discussion}

\end{document}